\documentclass[twocolumn,amsmath,floats,showpacs,nofootinbib]{revtex4}
\usepackage[usenames]{color}
\usepackage{graphicx}% Include figure files
\usepackage{epstopdf}
\usepackage{textcomp}
\usepackage{hyperref}
\usepackage{multirow}
\usepackage{tikz}
\usepackage{rotating}

\hypersetup{colorlinks=true}

\begin{document}

\title{Point-contact spectroscopy of the high-temperature superconductor BiSrCaCuO}

\author{L.F. Rybal'chenko, V.V. Fisun, N.L. Bobrov, M.B. Kosmyna, A.I. Moshkov, V.P. Seminozhenko, and I.K. Yanson}
\affiliation{Physicotechnical Institute of Low Temperatures, Academy of Sciences of the Ukrainian SSR, Kharkov,\\
A. M. Gorky State University, Kharkov,\\
and Scientific Industrial Complex "Monokristallreaktiv", Kharkov\\
Email address: bobrov@ilt.kharkov.ua}
\published {(\href{http://fntr.ilt.kharkov.ua/fnt/pdf/15/15-1/f15-0095r.pdf}{Fiz. Nizk. Temp.} , \textbf{15}, 95 (1989)); (Sov. J. Low Temp. Phys., \textbf{15}, 54 (1989)}
\date{\today}

\begin{abstract}The maximum value of the energy gap $\Delta\simeq 8\ meV$ and the ratio $2\delta/kT_c\simeq~2.5$ are determined for a high-temperature superconductor $\rm Bi_2Sr_2CaCu_2O_{8+y}$ by using point contacts. It was found that the high-temperature superconductor is transformed under the effect of current injection of quasiparticles to a new modified state with a reduced gap, which is stable in a wide range of injection intensity.

\pacs { 74.45.+c, 73.40.-c, 74.20.Mn, 74.72.-h, 74.72.Jt}

\end{abstract}

\maketitle

A new class of high-temperature metal oxide superconductors BiSrCaCuO, synthesized recently\cite{1} is distinguished from the previous types of HTS (LaSrCuO and YBaCuO) by a number of peculiar properties, such as high temperatures of the onset of the superconducting transition $T_{c0} \simeq 120~K$, the clearly planar type of the Cu-O bonds, and the absence on one-dimensional $\rm Cu-O$ chains of the vacancy type in the layers, and slightly higher resistivities. This makes the BiSrCaCuO system extraordinary and stimulates the investigation into its physical properties, above all, the energy gap in the quasiparticle excitation spectrum. It is also aimed at a deeper understanding of the nature of HTS. An attempt to measure the energy gap with the help of the tunnel effect was made in one of the first works in this direction \cite{2}, but no positive results were obtained.

Here we report on the results of investigation of polycrystalline samples of $\rm Bi_2Sr_2CaCu_2O_{8+y}$ by the point-contact (PC) spectroscopy method, which was found to be fruitful in studying the LaSrCuO and YBaCuO compounds \cite{3,4,5,6}. Data were obtained on the value of the energy gap. It was found that processes similar to the formation of phase slip centers (PSC) in narrow superconducting channels occur in point contacts with an undeformed constriction region under the effect of current injection. It was found that for a high injection intensity HTS goes over to a new modified state with a reduced gap.
\begin{figure}[]
\includegraphics[width=8cm,angle=0]{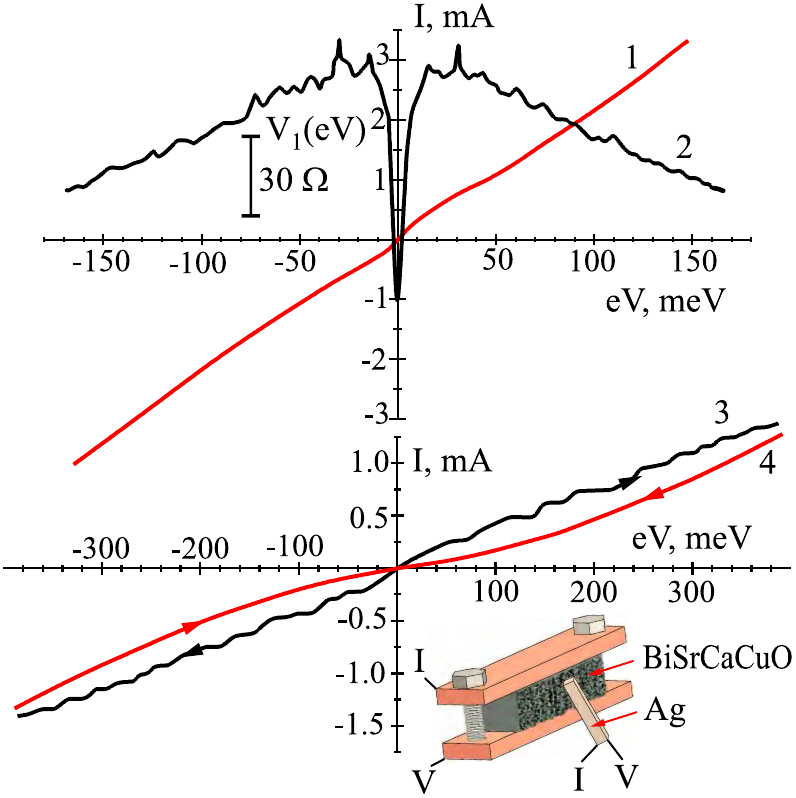}
\caption[]{IVC (1) and its first derivative $V_1(eV)\sim dV/dI$ (2) at 15~$K$ for the BiSrCaCuO-Ag heterojunction with $R_0\simeq 160~\Omega$ (for $V=0$) and the emergence of hysteresis in IVC in the crystallographically ordered BiSrCaCuO-Ag contact ($R_0\simeq 150~\Omega, T\simeq 4.2~K$) after applying high bias voltages ($V>500~mV$). Direct (3) and reverse (4) branches of IVC coincide for such a bias voltage.}
\label{Fig1}
\end{figure}
Point contacts of the ScS and ScN type (S stands for BiSrCaCuO, ñ for constriction, and N for Ag) of submicrometer size were formed using the displacement technique \cite{7} between contacting sharp edges of corresponding electrodes (see inset to Fig. \ref{Fig1}). S-electrodes with a size of several millimeters were cleaved from bulk polycrystalline HTS samples. The latter were prepared by the solution-melt technique with copper compounds as a solvent. According to resistive measurements, the superconducting transition in bulk samples started at 120~$K$ and terminated at about 82~$K$. Point contacts of direct conductivity type were chosen for analysis. They are characterized by the presence of clearly manifested excess current $I_{exc}$ in current-voltage characteristics (IVC).

Both IVC and their derivatives $V_1(eV)\sim dV/dI$, carrying information about the energy gap in the superconducting electrode were recorded in the process of measurement. Typical dependences for the BiSrCaCuO-Ag heterojunction are shown in Fig. \ref{Fig1} (curves 1 and 2). Unlike LaSrCuO and especially YBaCuO, a nonlinearity due to superconductivity appeared practically at each point of contact between Ag and the BiSrCaCuO surface being probed. However, the magnitude of excess current was usually small. The structureless shape of the central minimum (in the vicinity of $V=0$) on the $V_1(eV)$ dependence (see Fig. \ref{Fig1}, curve 2) can be associated either with a complete absence of the energy gap in the quasiparticle excitation spectrum, or with its
small value ($\Delta < 3~meV$). It can also be due to the presence on the surface of grains of thin (of the order of coherence length $\xi$) layers of S with considerably suppressed superconducting
parameters. The latter factor looks most probable on account of the experience gained in operating with point contacts made of LaSrCuO and YBaCuO \cite{3,5}.

For a number of fused HTS samples, the IVC of corresponding contacts exhibit a clear step structure (see Fig. \ref{Fig1}, curve 3) which was observed earlier in contacts made of YBaSrCuO single crystals with a high degree of crystallographic ordering in the constriction \cite{8}. This structure was attributed to the discrete nature of penetration of electric field into the point-contact constriction region, i.e., to the formation of a specific spatial structure, viz., phase slip surfaces similar to those observed in narrow superconducting channels \cite{9}.

\begin{figure}[]
\includegraphics[width=8cm,angle=0]{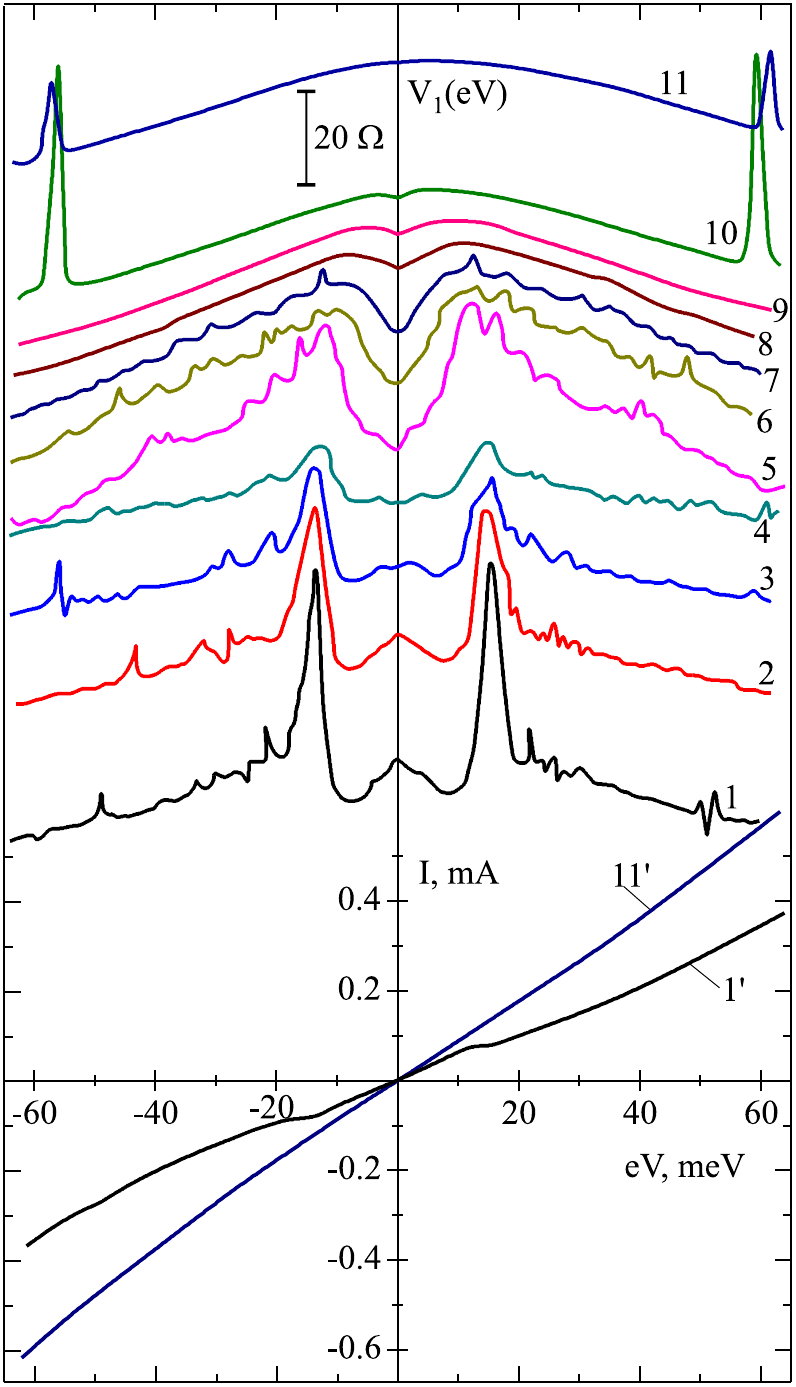}
\caption[]{IVC (l',11') and $V_1(eV)$ dependences (1-11) for a homojunction made of BiSrCaCuO at $T$=4.2 (1,1'), 7(2), 10(3), 20(4), 30(5), 40(6), 50(7), 60(8), 65(9), 70(10), and 75 (11,11')~$K$; $R_0(4.2~K)\simeq140~\Omega$, $R_0(75~K)\simeq101~\Omega$.}
\label{Fig2}
\end{figure}

A large number of current steps observed in IVC (Fig. \ref{Fig1}, curve 3) indicates that the point-con- tact diameter $d\gg\xi(\xi\sim$30~\AA) \cite{1} and is at least several hundred angstroms in the order of magnitude. In the case under consideration, it is difficult to estimate the value of d from the contact resistance $R_{con}$, say, in diffusion mode, by using Maxwell's formula $d=\rho_{con}/R_{con}$ ($\rho_{con}$ is the resistivity of the substance in the point-contact constriction region). This is due to the strong anisotropy of resistivity of the substance under investigation ($\rho_c/\rho_{ab}\sim 10^{4}-10^{5}$) \cite{10} and the indeterminacy of the crystallographic orientation of the point contact. Therefore, taking for $\rho_{con}$ the value corresponding to the direction parallel to the basal plane ($\rho_{ab}\sim10^{-4}~\Omega\cdot cm$), we can obtain only the lower limit $d_{min}$. For typical values of $R_{con}\sim10^2~\Omega$, we obtain $d_{min}=10^2$~\AA. The real value of $d$ can be considerably higher not only because of the noncoincidence between the contact axis and the basal plane of the HTS, but also due to the fact that a nonsuperconducting phase of the semiconductor type may get into the constriction region. The low values of excess currents and the activation nature of the point-contact conductivity are indications of this effect.

The existence of a continuous layer of S' with degraded superconductivity on all cleaved surfaces of BiSrCaCuO under investigation, which cannot be removed by the displacement of electrodes, is confirmed by the absence of the steady-state Josephson effect in the geometry of homojunctions (ScS).

Gap minima (Fig. \ref{Fig2}) were observed on the $V_1(eV)$ dependences in the region of increasing $I_{exc}$ for many homojunctions. This allowed us to determine the energy gap and trace its temperature dependence. It was found that $\Delta\rightarrow 0$ at temperatures that are several tens of degrees lower than $T_c^* (T_c^* \sim 75-80~K$ was determined from PC characteristics ). The value $\Delta\simeq 8~meV$ corresponding to the ratio $2\Delta/kT_c^* \simeq 2.5$ was recorded for a number of contacts. In some cases the values of $\Delta \simeq 10-11~meV$ were obtained. Two gap minima corresponding to $\Delta_1\simeq 8~meV$ and $\Delta_2\simeq 3~meV$ were recorded for several homojunctions in the region of increasing $I_{exc}$. In all probability, the S' layer is also responsible for lower values of $\Delta$ for LaSrCuO and YBaCuO in comparison with those for $\Delta$ and BiSrCaCuO. In the case of homojunctions, it was possible to make it very thin due to the high hardness of the two electrodes. It should be noted that weak gap minima on the $V_1(eV)$ dependences were also observed for some heterojunctions BiSrCaCuO-Ag in the vicinity of 5~$meV$.

\begin{figure}[]
\includegraphics[width=8cm,angle=0]{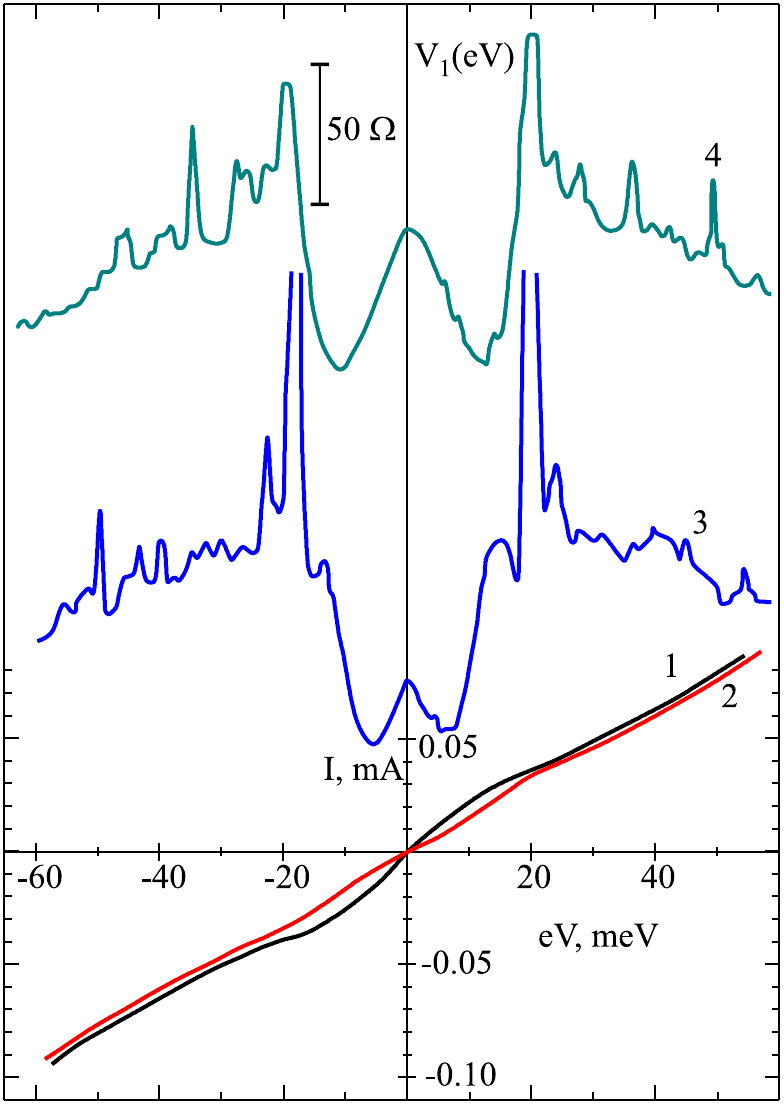}
\caption[]{The effect of microwave (7.4 GHz) power P on IVC (1,2) and dependence $V_1(eV)$ (3,4) for BiSrCaCuO homojunction at $T\simeq4.2~K$,\ P=0(1,3), and $\rm P=P_{max}$ (2,4), $R_0\simeq300~\Omega$.}
\label{Fig3}
\end{figure}

The presence of the defective S' layer on the $\rm BiSrCaCuO$ surface affects the behavior of the point contact in the microwave field. Figure \ref{Fig3} reflects the influence of the microwave radiation on IVC and the $V_1(eV)$ dependence of one of the homojunctions. The decrease in the value of $I_{exc}$ caused by microwave field (curves 1 and 2) is mainly limited to the region of its increase and is reduced insignificantly in the higher-energy region ($eV\gg\Delta$). This effect is accompanied by a shift in the gap minima on  $V_1(eV)$ toward higher energies (curves 3 and 4).
The effect of the microwave field can be explained by an increase in the sharpness of the S'-S boundary due to the complete suppression of the gap in the S' layer. The observed (increased) gap in the S layer is in a better agreement with its deep unperturbed value ( $\Delta\simeq 12~meV$) than in the absence of the microwave field when the recorded gap is smaller than its true value due to the proximity effect in S'-S. The emergence of an additional potential drop in the S' layer can make a certain contribution to the observed shift of the gap minima. An
insignificant decrease in the value of $I_{exc}$ the region of high bias voltage is due to the fact that the excess current here is mainly caused by Andreev reflection of quasiparticles from the regions situated far from the center of the junction, and hence less affected by the microwave radiation. It should be noted that in the region of high bias voltage and without a microwave field, the superconductivity in the S' layer can be suppressed by the high intensity of the current injection of quasiparticles.

An unusual phenomenon observed for most of crystallographically ordered contacts with a steplike IVC is worth mentioning. We speak about the appearance of a noticeable hysteresis on IVC recorded with the direct and reverse current scanning (or vice versa) (see Fig. \ref{Fig1}, curves 3 and 4). In many cases, the hysteresis embraces nearly the entire interval of bias voltages applied to the contact without any noticeable phase slip effect on the reverse branch of IVC. Such a situation emerges, as a rule, after high bias voltages are obtained ($\sim 500~mV$ or higher). These voltages cannot be withstood by every contact. It should be emphasized that in repeated recordings in the forward direction (starting with $I, V = 0$), the step structure of IVC is always reproduced.

This phenomenon cannot be explained by a trivial Joule heating of PC region due to an increase in the critical current density in superconducting channels. Indeed, an increase in temperature normally leads to a decrease in the resistance of the contact (Fig. \ref{Fig2}, curves 1' and 11'), while the reverse hysteresis branch in IVC corresponds to its increase.

The reason behind such a deep hysteresis can be associated with peculiarities of the nonequilibrium state which takes place in the point-contact constriction region under the effect of the current injection of quasiparticles into the superconductor. One aspect of this state, which is determined by short lengths of conversion of quasiparticles into pairs in HTS, is manifested in the point contact in the form of current steps on IVC. There is, however, another aspect of the nonequilibrium problem, which has not been studied in detail. An analysis of the
effect of the current injection on the superconductor, carried out earlier in Ref. \cite{11}, led to two-valued solutions for the gap parameter in a wide range of injection intensity. Hence it cannot be ruled out that the reverse hysteresis branch recorded in our experiments corresponds to lower values of $\Delta$ which for some reason may become stable after attaining the critical power of injection of nonequilibrium quasiparticles. Therefore, a transition to a new modified state of HTS, which is stable in the wide range of the current injection intensity of quasiparticles, can be made.

From this point of view, the absence of current steps in the reverse hysteresis branch in IVC becomes clear. As a matter of fact, the length $l_E$ over which injected quasiparticles are transformed into pairs is determined to a considerable extent by the time of charge imbalance relaxation $\tau_Q\sim\Delta^{-1}$. The value of $\tau_Q$ increases with decreasing $\Delta$, and hence $l_E$ also increases, since $l_E\propto\tau_Q$. As a result, $l_E$ may exceed $d$, and the phase slip effects in point contact may vanish.

Another explanation of the hysteresis can be associated with an accumulation of a space charge in the S' layer (of a semiconducting nature) followed by its gradual dissipation. The parabolic shape of IVC points toward the accumulation of space charge and the emergence of currents bounded by the space charge without clear indications of a superconducting gap on IVC \cite{12}.

\vspace{\baselineskip}
\textbf{NOTATION}\vspace{\baselineskip}
\\
Here $T_{c0}$ is the temperature of the onset of the superconducting transition, $I_{exc}$ the excess current, $\Delta$ the superconducting energy gap, $\xi$ the coherence length, $\rho_{con}$ the resistivity of the substance in the region of point-contact constriction, $\rho_c$ and $\rho_{ab}$ the resistivity of HTS along the $ñ$ axis and in the $ab$ plane, \emph{d} the point-contact diameter, $T_c^*$ the critical temperature of the superconducting transition onset in point contact, $l_E$ the length over which injected quasiparticles are transformed into pairs, and $\tau_Q$ the charge imbalance relaxation time.

\end{document}